\documentclass[%
 reprint,
 amsmath,amssymb,superscriptaddress,
 aps,
]{revtex4-2}

\usepackage[dvipsnames]{xcolor}

\usepackage{graphicx}
\usepackage{dcolumn}
\usepackage{bm}

\begin{document}
\title{Uniaxial strain control of bulk ferromagnetism in rare-earth titanates}

\author{A. Najev}
\affiliation{Department of Physics, Faculty of Science, University of Zagreb, Bijeni\v cka 32, HR-10000 Zagreb, Croatia}
\affiliation{School of Physics and Astronomy, University of Minnesota, Minneapolis, MN 55455, U.S.A.}
\author{S. Hameed}
\affiliation{School of Physics and Astronomy, University of Minnesota, Minneapolis, MN 55455, U.S.A.}
\author{D. Gautreau}
\affiliation{School of Physics and Astronomy, University of Minnesota, Minneapolis, MN 55455, U.S.A.}
\affiliation{Department of Chemical Engineering and Materials Science, University of Minnesota, Minneapolis, MN 55455, U.S.A.}
\author{Z. Wang}
\affiliation{School of Physics and Astronomy, University of Minnesota, Minneapolis, MN 55455, U.S.A.}
\author{J. Joe}
\affiliation{School of Physics and Astronomy, University of Minnesota, Minneapolis, MN 55455, U.S.A.}
\author{M. Po\v{z}ek}
\affiliation{Department of Physics, Faculty of Science, University of Zagreb, Bijeni\v cka 32, HR-10000 Zagreb, Croatia}
\author{T. Birol}
\affiliation{Department of Chemical Engineering and Materials Science, University of Minnesota, Minneapolis, MN 55455, U.S.A.}
\author{R. M. Fernandes}
\affiliation{School of Physics and Astronomy, University of Minnesota, Minneapolis, MN 55455, U.S.A.}
\author{M. Greven}
\affiliation{School of Physics and Astronomy, University of Minnesota, Minneapolis, MN 55455, U.S.A.}
\author{D. Pelc}
\affiliation{Department of Physics, Faculty of Science, University of Zagreb, Bijeni\v cka 32, HR-10000 Zagreb, Croatia}
\affiliation{School of Physics and Astronomy, University of Minnesota, Minneapolis, MN 55455, U.S.A.}

\begin{abstract} 
The perovskite rare-earth titanates are model Mott insulators with magnetic ground states that are very sensitive to structural distortions. These distortions couple strongly to the orbital degrees of freedom and, in principle, it should be possible to tune the superexchange and the magnetic transition with strain. 
We investigate the representative system (Y,La,Ca)TiO$_3$, which exhibits low crystallographic symmetry and no structural instabilities. From magnetic susceptibility measurements of the Curie temperature, we demonstrate direct, reversible and continuous control of ferromagnetism by influencing the TiO$_6$ octahedral tilts and rotations with uniaxial strain. 
The relative change in $T_C$ as a function of strain is well described by \textit{ab initio} calculations, which provides detailed understanding of the complex interactions among structural, orbital and magnetic properties in rare-earth titanates. The demonstrated manipulation of octahedral distortions opens up far-reaching possibilities for investigations of electron-lattice coupling, competing ground states, and magnetic quantum phase transitions in a wide range of quantum materials.
\end{abstract}
\maketitle

Transition-metal oxides exhibit a wealth of distinct structural, magnetic and orbital ordering tendencies, and are thus among the most extensively studied condensed matter systems. A salient feature of these materials is an intricate interplay between structural and electronic properties: ferroelectric, metal-insulator and superconducting transitions can all be strongly influenced by changes in the crystalline lattice. Importantly, the electronic ground states can thus be tuned by manipulating the structure. This has been employed in a wide range of materials, and led to significant breakthroughs, with the emergence of uniaxial strain as a particularly interesting control variable \citep{haeni,takeshita2004,hardy2010,fisher2010,hicks2,hicks22,hicks4,vana,vana2,hicks3,pnictides}. Prominent examples include the stabilization of ferroelectricity in epitaxially-strained films of strontium titanate \cite{haeni}, as well as uniaxial stress manipulation of superconductivity in strontium ruthenate \cite{hicks2, hicks22, hicks4}, metal-insulator transitions in vanadium oxides \cite{vana, vana2}, superconductivity and charge-density-wave order in cuprates \cite{takeshita2004,hicks3, uniax_cuprate}, and antiferromagnetism in pnictides \cite{pnictides}. However, this approach has so far relied either on the proximity to a structural instability, or on the strain-induced lowering of structural symmetry, and it has not yet been applied to bulk ferromagnetic (FM) materials. 

\begin{figure}
\includegraphics[width=0.48\textwidth]{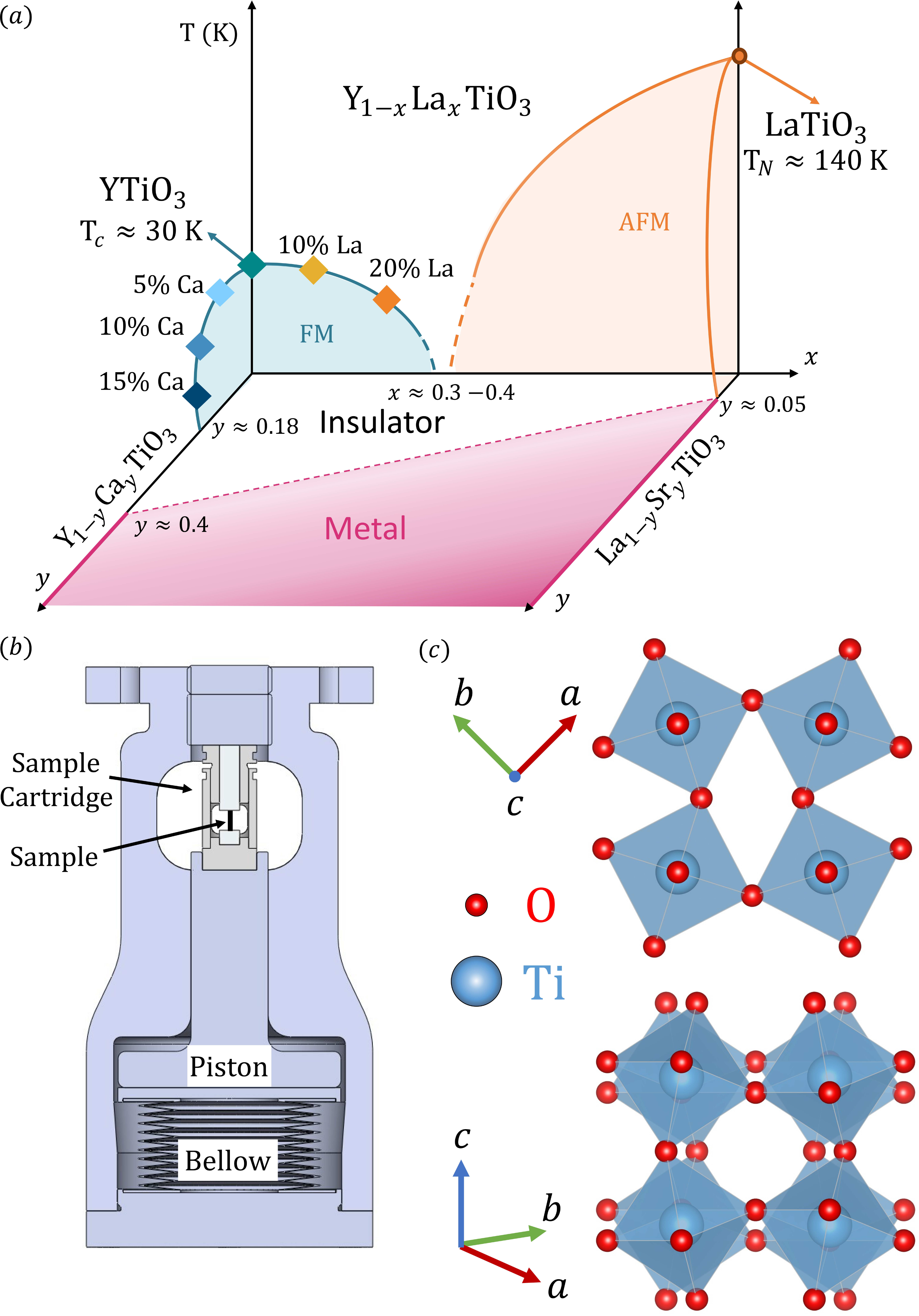}
\caption{\label{fig:LTcelija} (color online) (a) Phase diagram of RTiO$_3$ with FM (ferromagnetic), AFM (antiferromagnetic), insulating, and metallic phases. The compositions studied in this work are marked by diamonds. (b) Schematic of the uniaxial strain cell (for more details, see \cite{Note1}). (c) YTiO$_3$ structure constructed from theory, showing the two relevant octahedral rotations, $a^0a^0c^+$ and $a^-a^-c^0$, as deviations from a simple cubic structure. Small spheres depict oxygen ions, large spheres titanium ions. Upper figure: oxygen base plane with the $a^0a^0c^+$ rotation mode. Bottom figure: perpendicular view showing the $a^-a^-c^0$ rotation mode. The effects of uniaxial stress along $a \equiv [100]$, $b \equiv [010]$ or $c \equiv [001]$ on the rotation modes are qualitatively different for each direction.}
\end{figure}

The trivalent rare-earth (RE) titanates RTiO$_3$ (R is a rare-earth ion) are prototypical three-dimensional Mott insulators \cite{revijalac} with rich phase diagrams (Fig.\,\ref{fig:LTcelija}(a)) that are not fully understood. 
The R ion is surrounded by eight TiO$_6$ octahedra \cite{struktura}, but due to a considerable atomic-size mismatch, the lattice symmetry is orthorhombic (\textit{Pbnm}), significantly lower than the ideal cubic perovskite. 
The TiO$_6$ octahedra are both tilted and rotated, and the distortions are more pronounced for smaller R ions, leading to larger deviations of the Ti-O-Ti bond-angle from 180$^\circ$ \cite{struktura} (Fig.\,\ref{fig:LTcelija}(c)). 
Since the orbital overlap strongly depends on the bond angles, so does the superexchange interaction between unpaired electron spins associated with the Ti $3d$ t$_{2g}$ orbitals. Consequently, the spin-lattice coupling is strong \cite{knafo,hem,magnetoelast}, and the magnetism in RE titanates can be tuned by varying the average R-ion size: the magnetic ground state changes from FM to antiferromagnetic (AFM) with increasing R-ion size \cite{gree,katsu}, or upon atomic substitution, {\it e.g.}, in Y$_{1-x}$La$_x$TiO$_3$ (YLTO) \cite{oki,goral} (Fig.\,1(a)). Charge doping can also be used to control the magnetic ground state, \textit{e.g.}, in Y$_{1-y}$Ca$_y$TiO$_3$ (YCTO), where hole doping destroys the long-range FM order at $y \sim 20\%$ and eventually leads to an insulator-metal transition \cite{tokura,HameedYCa2021}. Based on the behavior of thermal expansion coefficients across the magnetic transitions, it has been suggested that uniaxial strain might have an effect similar to atomic substitution; strain along the orthorhombic $a$ direction should increase the octahedral distortions, stabilize the FM order, and increase the Curie temperature ($T_C$), whereas strain along $b$ should decrease the distortion and $T_C$, similar to the substitution of Y with La (Fig.\,1(c)) \cite{knafo}. Attempts to control $T_C$ with hydrostatic pressure were previously made for pressures up to 800\,MPa \cite{krit}. However, the results show only a slight decrease of $T_C$. 

In this Letter, we show that the Curie temperature of the YTO-based materials can be manipulated via \textit{in situ} uniaxial stress in a remarkably wide range. We study both La- and Ca-substituted YTO, in the substitution range in which a FM ground state is observed (Fig.\,1(a)), and demonstrate that $T_C$ can be reversibly and continuously suppressed or enhanced by up to a factor of $\sim 2$, depending on the specific crystalline direction in which the uniaxial stress is applied. Moreover, we obtain nearly complete suppression of ferromagnetism in a Ca-substituted sample close to the FM-paramagnetic phase boundary. 
Through a comparison with \textit{ab initio} and mean-field calculations, we show that the origin of the observed behavior is the strong effect of stress on the octahedral rotation distortions, and the pronounced sensitivity of the magnetic exchange couplings on these distortions. We thus demonstrate the potential of uniaxial stress engineering of oxygen octahedral rotations, which are present in most perovskite-based oxides, as a practical means to manipulate magnetism and induce quantum phase transitions.

We employ a uniaxial pressure-cell design \footnote{See Supplemental Material at url for details on the experimental setup and the first-principles calculations, with additional references \cite{kresse1993, kresse1996CMS, kresse1996PRB, perdew2008, dudarev1998, stokes2006}}\setcounter{footnote}{16} (Fig.\,\ref{fig:LTcelija}(b)) that enables the application of high and spatially homogeneous uniaxial stress on millimeter-sized single crystals \citep{pelc2019,hameed2020}. Force on the sample is generated with a helium pneumatic system, with bellows that expand when filled with pressurized gas and push on a piston that compresses the sample. By controlling the gas pressure, the force can be finely tuned. Samples are mounted within a cartridge with an inductive displacement sensor that allows independent determination of sample stress and strain.
The strain-dependent magnetic susceptibility is measured \textit{in situ} with either a mutual inductance setup or an inductance bridge \cite{Note1}. The samples are single crystals of YTiO$_3$, YLTO and YCTO, grown by the traveling-solvent floating-zone method \cite{HameedYLa2021,HameedYCa2021, HameedGrowth}, that were extensively characterized with Laue x-ray diffraction and polished to high precision along the orthorhombic directions $a$, $b$ and $c$.

\begin{figure}
\includegraphics[width=8.6 cm]{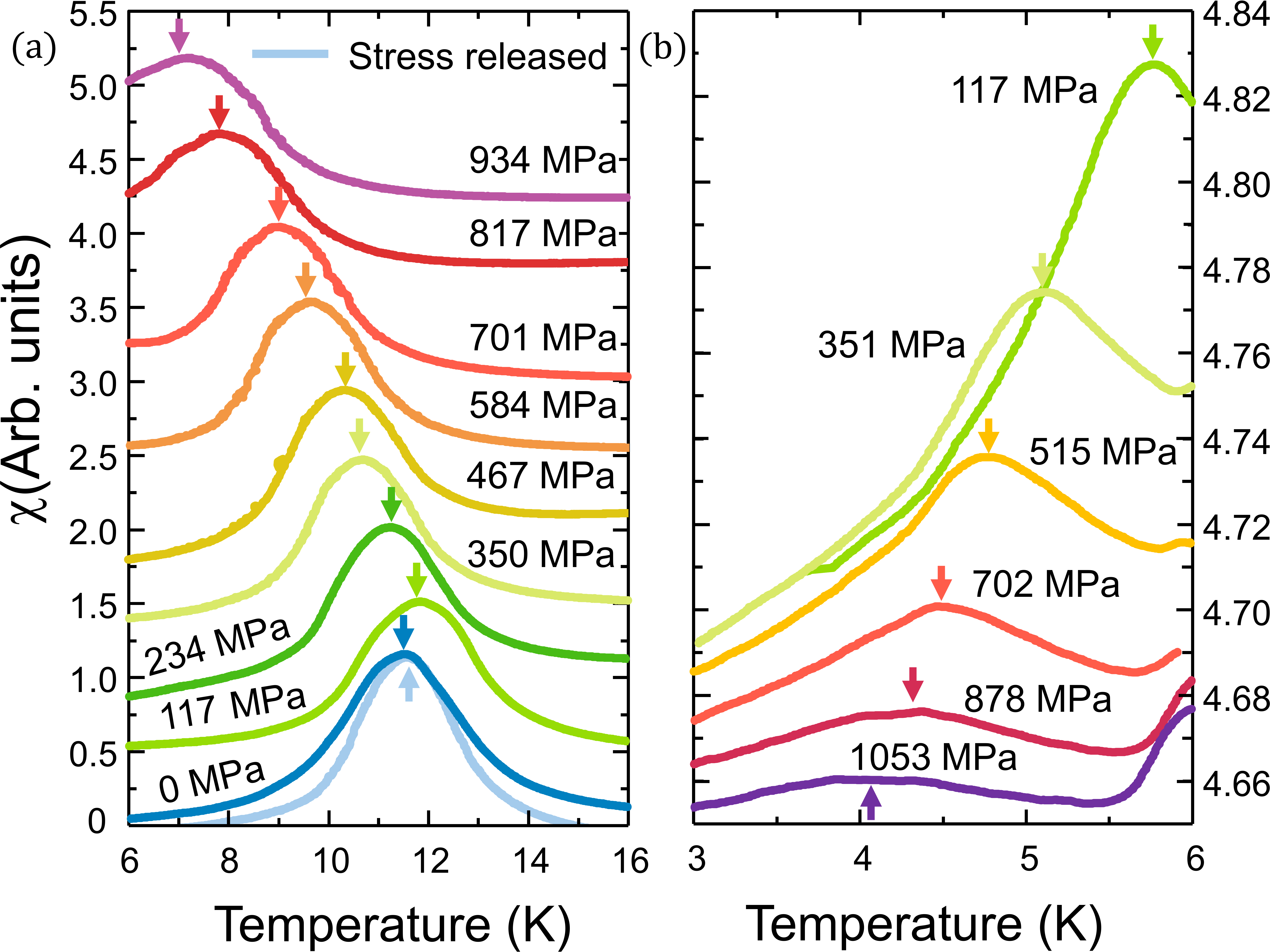}
\caption{\label{fig:raw} (color online) Raw ac susceptibility data for (a) Y$_{0.9}$Ca$_{0.1}$TiO$_3$ and (b) Y$_{0.85}$Ca$_{0.15}$TiO$_3$, with applied stress along $b$ and susceptibility measured along $c$. The Curie temperatures are taken as the peak positions (arrows). Data for different sample stresses are shifted vertically for clarity. In (a), the negligible increase of the susceptibility peak width with increasing stress indicates that the strain is homogeneously distributed in the sample; see \cite{Note1} for the extracted full-width-at-half-maximum values. After stress release, $T_C$ returns to its original value; the peak narrows slightly, indicative of residual inhomogeneous stress in the samples that decreases after the application of external pressure. In (b), the FM order disappears at a nonzero temperature, consistent with a first-order transition into a paramagnetic state. The small peak around 6\,K likely originates from superconducting solder in the detection coils \cite{Note1}.}
\end{figure}

The nonmagnetic symmetry of the RE titanate perovskites is low enough to preclude collinear magnetic order. As a result, spin canting is present in both FM and AFM phases, which have the same magnetic space group \cite{magnetoelast}. The FM moments in, \textit{e.g.}, YTiO$_3$ order predominantly along $c$, whereas the AFM moment in LaTiO$_3$ is predominantly of G-type and orders along $a$ \cite{ulrich2002}. No long-range structural changes are known to occur around the Curie and N\'eel temperatures, but due to the strong and anisotropic spin-lattice coupling, we observe significant effects of uniaxial stress on $T_C$. Representative ac susceptibility measurements for two samples, Y$_{0.9}$Ca$_{0.1}$TiO$_3$ and Y$_{0.85}$Ca$_{0.15}$TiO$_3$, are shown in Figs. \ref{fig:raw}(a) and (b), respectively. Both were deformed along $b$, which leads to a substantial decrease of $T_C$. The deformation is also seen to be reversible, despite the large stress values, as $T_C$ before and after deformation is essentially the same. Moreover, the susceptibility curves and transition temperatures obtained for different stress levels are quite smooth, suggesting that no structural transitions occur with applied stress \cite{Note1}. The Ca-15\% sample is close to the boundary of FM order in the phase diagram of YCTO, and we were able to suppress the magnetism almost completely with strain. Interestingly, $T_C$ appears to level off at high strain, but the magnitude of the susceptibility peak decreases toward zero. A possible explanation for this is that the FM volume fraction gradually decreases to zero with increasing uniaxial strain, thereby reducing the average bulk ordered moment. This would be indicative of a first- rather than a second-order transition between the FM and paramagnetic phases, with the associated phase separation influenced by strain. 

\begin{figure*}
\includegraphics[width=175mm]{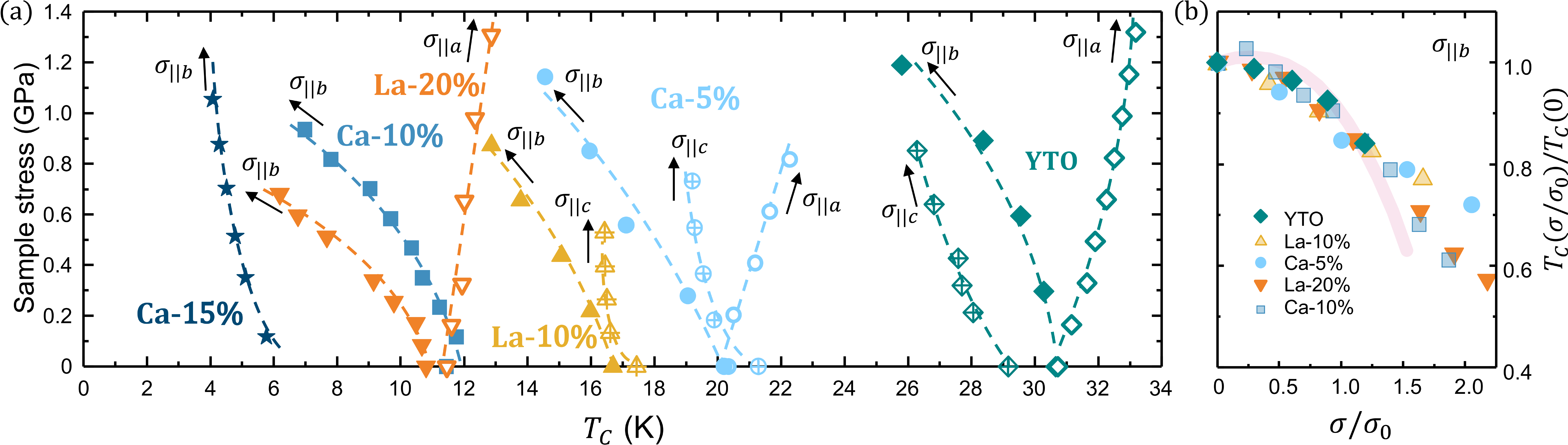}
\caption{\label{fig:resfd} (color online) (a) Phase diagram of $T_C$ \textit{vs.} applied sample stress, spanning a wide range of $T_C$ values. Empty, filled, and crossed symbols: applied stress along $a$, $b$, and $c$, respectively (marked as $\sigma_{||a,b,c}$). Green symbols: YTiO$_3$ (YTO). Light blue, blue, and dark blue: Y$_{0.95}$Ca$_{0.05}$TiO$_3$, Y$_{0.9}$Ca$_{0.1}$TiO$_3$, Y$_{0.85}$Ca$_{0.15}$TiO$_3$, respectively. Yellow and orange: Y$_{0.9}$La$_{0.1}$TiO$_3$ and Y$_{0.8}$La$_{0.2}$TiO$_3$, respectively. Lines are guides to the eye. Small differences between the $\sigma=0$ transition temperatures for the same nominal compositions are due to a slight variation in oxygen off-stoichiometry of crystals from the same growths \cite{HameedGrowth}. (b) Single-parameter scaling of $T_C(\sigma/\sigma_0)/T_C(0)$ with stress along $b$. The doping-dependent scaling parameter $\sigma_0$ plays the role of an effective elastic modulus. Line: DFT calculation for YTiO$_3$ (same as in Fig. 4c). See \cite{Note1} for values of $\sigma_0$.}
\end{figure*}

Figure \ref{fig:resfd} summarizes the Curie temperatures of all measured samples in a stress \textit{vs.} $T_C$ phase diagram. Our results directly confirm the proposed qualitative phase diagram of ref. \cite{knafo}: Stress along $b$ and $c$ decreases $T_C$, whereas compression along $a$ promotes ferromagnetism and increases $T_C$. 
Our measurements reveal a sizable nonlinearity of $T_C$ at high stress values and, as noted, a leveling off of $T_C$ as the FM-paramagnetic phase boundary is approached with increasing stress in Y$_{0.85}$C$_{0.15}$TO$_3$. Although the influence of pressure on $T_C$ qualitatively agrees with the predicted phase diagram, we note that significant irreversibility appears for stress along $c$, even at stress levels as low as 100\,MPa, with $T_C$ decreasing only slightly after an initial shift (see \cite{Note1} for details). This effect is most prominent for the doped samples and might be related to intrinsic structural inhomogeneity. In contrast, when stress is applied along $a$ and $b$, the strain (and therefore the change in $T_C$) is highly reversible. 

With the exception of Y$_{0.85}$C$_{0.15}$TO$_4$, the dependence of $T_C$ on stress is qualitatively similar for all studied samples. To perform a quantitative comparison, we introduce the scaling parameter $\sigma_0$, an effective elastic modulus that connects applied stress and octahedral distortions. We take the reference value $\sigma_0 = 1$~GPa for YTiO$_3$, and then adjust $\sigma_0$ for doped samples to obtain the scaling shown in Fig.\,3(b) (see \cite{Note1} for $\sigma_0$ values). All data for stress along $b$ (except for Ca-15\%, as noted) collapse on the master curve, with the doping-dependent $\sigma_0$ as the only free parameters. This suggests that neither Ca- nor La-doping fundamentally changes the physics of spin-lattice coupling, at least for doping levels not too close to the FM-paramagnetic boundary. Instead, doping mainly affects the mechanical properties: $\sigma_0$ increases with doping \cite{Note1}, implying that the lattice is more susceptible to tilts for a given applied stress.

\begin{figure*}
\includegraphics[width=\textwidth]{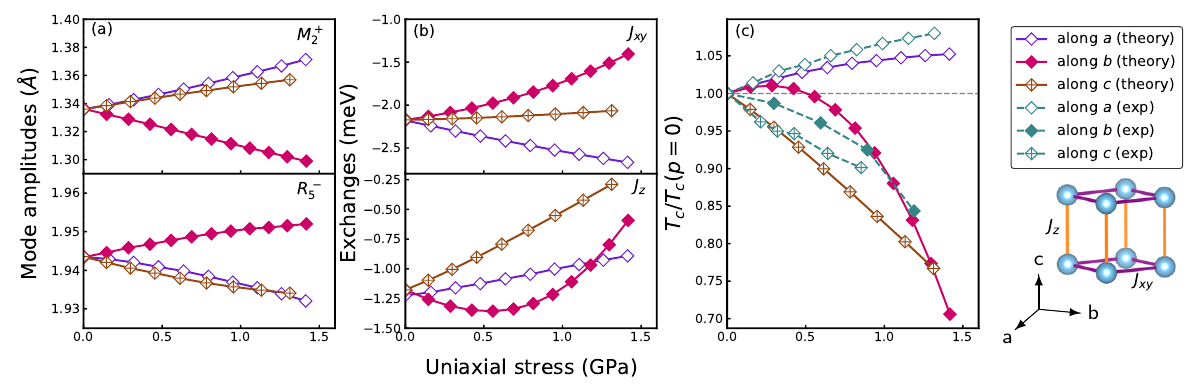}
\caption{\label{fig:theory} (color online) DFT results for YTiO$_3$ as a function of compressive uniaxial stress along $a$, $b$, and $c$. (a) Mode amplitudes of the octahedral rotations corresponding to the $M_2^+$ and $R_5^-$ irreps in YTiO$_3$. (b) Nearest-neighbor exchange parameters $J_{xy}$ and $J_z$ obtained from fits to the Heisenberg model. (c) Mean-field Curie temperatures using the DFT-obtained exchanges, including the sub-leading next-nearest-neighbor exchanges \cite{Note1}, along with experimental results for YTiO$_3$. A similar analysis for tensile strain can be found in \cite{Note1}.
}
\end{figure*}

In order to model the observed stress dependence of $T_C$ and to elucidate the connection between the crystal structure and magnetic interactions, we performed first-principles DFT calculations for YTiO$_3$ to predict the crystal structure and the exchange parameters for a Heisenberg model under uniaxial stress along each of the orthorhombic principal axes \cite{Note1}. Calculations for doped systems were not performed due to the exceedingly high numerical cost, but given the scaling in Fig.\,3(b), the results for the undoped system should also apply to doped compounds.
Note that previous studies have shown the suitability of a simple nearest-neighbor Heisenberg model to capture most of the spin-wave dispersion properties \cite{ulrich2002}. Here, our main modifications are to allow for the exchange interaction to differ along in-plane and out-of-plane directions, \textit{i.e.} $J_{xy} \neq J_z$ (see inset of Fig.\,\ref{fig:theory}) and to include sub-leading next-nearest-neighbor interactions \cite{Note1}.

We focus on the two types of octahedral rotations in YTiO$_3$, namely the $a^0a^0c^+$ and $a^-a^-c^0$ rotation modes (Fig.\,\ref{fig:LTcelija}(c)) in Glazer notation, which transform, respectively, as the irreducible representations (irreps) $M_2^+$ and $R_5^-$ of the space group $Pm\bar{3}m$. We characterize the distortion magnitudes according to their irrep mode amplitudes, that may be understood as follows. Consider some general distortion of a crystal, described as a vector of atomic displacements. To determine the extent to which this distortion is composed of a particular octahedral rotation mode (corresponding to a particular irrep), we calculate the inner products of the total distortion vector with the irrep vectors. This then projects out the weight corresponding to the octahedral rotations, and the magnitude of the inner product is defined as the irrep mode amplitude. As one may expect, these mode amplitudes are proportional to the corresponding octahedral rotation modes, and for small distortions are also proportional to the rotation angles. However, angles can become ill-defined for sufficiently distorted structures, while the irrep mode amplitudes are always well-defined.

We find that the two types of rotations have opposite trends under uniaxial stress along different axes (Fig.\,\ref{fig:theory}(a)): the $M_2^+$ rotation angle decreases under compressive stress along $b$, but increases under compressive stress along $a$ and $c$. The magnitude of the $R_5^-$ distortions has the opposite trend, as it increases under compressive stress along the $b$ axis. Moreover, the amplitude of the $R_5^-$ distortions is less sensitive to uniaxial stress. Given the low structural symmetry and complex multi-tilt system, it is not straightforward to intuitively grasp the tilt behavior with applied stress; for a more detailed analysis, see \cite{Gautreau22}.

The changes in the octahedral rotations alter the Ti-O-Ti angles, which in turn significantly alters the magnetic exchange interactions \cite{goodenough1976, birol2012}, as shown in Fig.\,\ref{fig:theory}(b). For stress along $a$, both nearest-neighbor exchange interactions $J_{xy}$ and $J_z$ vary almost linearly with stress, with $J_{xy}$ increasing in magnitude, whereas the magnitude of $J_z$ decreases. We also find that $J_{xy}$ is more sensitive to stress than $J_z$. This, in addition to the larger number of $J_{xy}$ vs. $J_z$ bonds, results in an increase of mean field $T_C$ for stress along $a$, as shown in Fig.\,\ref{fig:theory}(c). 
For stress along $c$, $J_{xy}$ remains mostly unchanged while $J_z$ decreases in magnitude almost linearly, which results in a suppression of $T_C$ with stress (Fig.\,\ref{fig:theory}(c)).
Finally, for uniaxial stress along $b$, $|J_{xy}|$ decreases while $|J_z|$ first increases with stress. These two competing behaviors lead to a slight increase in $T_C$ for smaller stress values, as indeed also observed in experiment (Fig. 2(a)). Yet $T_C$ decreases rapidly under larger values of stress as $|J_z|$ also decreases. As shown in Fig.\,\ref{fig:theory}(c) and Fig.\,3(b), in all three cases the first-principles calculations agree well with the experimentally observed relative change in $T_C$. This shows that the effect of stress on the magnetic properties can be explained by the crystal structural changes, and that it is mediated by the octahedral rotations.

To conclude, we have shown that the FM transition in RE titanates can be continuously tuned using uniaxial stress, and we have obtained a microscopic understanding of the underlying coupling between structural distortions, orbital overlap, and effective spin interactions. The remarkable agreement between measured and calculated changes of $T_C$ with stress shows that the intricate octahedral distortions and their influence on the superexchange are well captured by our first-principles calculations. 
The strong and anisotropic stress response is robust across the doping/ionic substitution phase diagram and does not rely on a nearby structural instability, which implies that these and related materials have tremendous potential for strain engineering of magnetism, including through the use of epitaxial strain and/or in heterostructures. Indeed, first-principles calculations have predicted strong effects of lattice strain on magnetism in manganite and  vanadate oxides \cite{magnan_press, magnan_press2, V_press}, although experimental work is still scarce and mostly focused on epitaxial strain in thin films. Additionally, in cobaltates, hydrostatic pressure seems to have a significant effect on the bandwidth and magnetic order \cite{Co_press, Co_press2}. Thus, uniaxial stress could be an important parameter in exploring the complex nature of magnetic transition metal oxides.
Finally, our work opens the possibility to study the (quantum) phase transitions between the FM, AFM and paramagnetic ground states using highly homogeneous uniaxial stress as a control variable. The spin-lattice coupling for deformation along $b$ is large enough to induce such a transition in the Y$_{0.85}$Ca$_{0.15}$TiO$_3$ sample studied here, but our result indicates that the transition may not be continuous. A first-order transition and associated phase separation regime could be a generic feature of this hole-doped system, where recent experiments have shown that the metal-insulator transition likely also involves phase separation \cite{HameedYCa2021}. Yet more work is needed to clarify this issue, in particular uniaxial stress experiments on nominally stoichiometric materials that are closer to the phase boundaries than YTiO$_3$. In this regard, it has been recently proposed that uniaxial strain in isovalent chemically-substituted RE titanates could be used to induce a quantum critical end-point \cite{Zhentao2021}.

\acknowledgements

We thank M. Lukas and C. Leighton for fruitful comments and discussions. The work at the University of Minnesota was funded by the Department of Energy through the University of Minnesota Center for Quantum Materials, under Grant No. DE-SC-0016371. The work at the University of Zagreb was supported by the Croatian Science Foundation through Grant No. IP-01-2018-2970.


%

\end{document}